\documentclass[aps,prl,twocolumn,showpacs,superscriptaddress,groupedaddress]{revtex4}  
\usepackage{graphicx}  
\usepackage{dcolumn}   
\usepackage{bm}        
\usepackage{amssymb}   
\usepackage[latin1]{inputenc}
\usepackage{amsmath}
\usepackage{amsfonts}
\usepackage{amssymb}
\usepackage{setspace}
\usepackage{mathtools}
\usepackage{pstricks}
\usepackage{color}
\providecommand{\f}[2]{\frac{{#1}}{{#2}}}

\newcommand{\ee}[1]{\begin{equation}#1\end{equation}}
\newcommand{\ea}[1]{\begin{align}#1\end{align}}

\begin{document}

\title{Spacetime curvature and the Higgs stability during inflation}

\author{M.~Herranen}\email{herranen@nbi.ku.dk}
\affiliation{Niels Bohr International Academy and Discovery Center, Niels Bohr Institute,
University of Copenhagen, Blegdamsvej 17, 2100 Copenhagen, Denmark
}
\author{T.~Markkanen}\email{tommi.markkanen@helsinki.fi}
\affiliation{Helsinki Institute of Physics and Department of Physics, P. O. Box 64,
 FI-00014 University of Helsinki, Finland
}
\author{S.~Nurmi}\email{sami.nurmi@helsinki.fi}
\affiliation{Helsinki Institute of Physics and Department of Physics, P. O. Box 64,
 FI-00014 University of Helsinki, Finland
}
\author{A.~Rajantie}\email{a.rajantie@imperial.ac.uk}
\affiliation{Theoretical Physics, Blackett Laboratory, Imperial College, London SW7 2AZ, United Kingdom
}
\date{\today}

\begin{abstract}
It has been claimed that the electroweak vacuum may be unstable during inflation due to large
 fluctuations of order $H$ in case of a high inflationary scale as suggested by BICEP2.
We compute the Standard Model Higgs effective potential including UV-induced curvature
 corrections at one-loop level. We find that for a high
 inflationary scale a large curvature mass is generated due to RG running
 of non-minimal coupling $\xi$, which either stabilizes the potential against fluctuations
 for $\xi_{\rm EW} \gtrsim 6\cdot 10^{-2}$, or destabilizes it for
 $\xi_{\rm EW} \lesssim 2 \cdot 10^{-2}$ when the generated curvature mass is negative. Only in
 the narrow intermediate region the effect of the curvature mass may be significantly smaller.
\end{abstract}

\maketitle
After the confirmed detection of Standard Model (SM) Higgs a substantial
amount of work has been devoted to investigating its ramifications
in the early universe. The measured Higgs mass $m_{\rm H} \sim 125$ GeV
 lies in the range where no new physics between the electroweak scale and inflation
 is necessarily required by theoretical self-consistency
 \cite{higgs_vac,Degrassi:2012ry,Buttazzo:2013uya}.
 However, if primordial gravitational waves possibly suggested by BICEP2
 data \cite{Ade:2014xna}
  would be detected, the implied high inflationary
 scale $H\gg 10^{9}$ GeV has been argued to be in tension with the pure
 SM Higgs \cite{Espinosa:2007qp,higgsstability}.

According to the effective potential computed in Minkowski space the
SM vacuum is separated from the unstable false vacuum by a barrier
of height $V_{\rm max}^{1/4} \sim 10^{9}$ GeV
\cite{Degrassi:2012ry}. If $H\gg 10^{9}$ GeV, the inflationary
fluctuations of the effectively massless Higgs immediately bring the
field in the false vacuum as the probability density at the barrier
 scales as $P\sim {\rm exp}(-8\pi^2\,V_{\rm max}/3H^4)$
 \cite{Espinosa:2007qp,higgsstability}.
Within the Standard Model it would therefore appear rather unlikely
that our observable patch of the universe would have survived in the
SM vacuum for the observationally required $N\sim 60$ e-folds of
inflation. Stabilizing the SM vacuum against inflationary
fluctuations, $V_{\rm max}^{1/4}\gtrsim H$, would require either a
low top mass at least two sigma below the best fit value or new
physics modifying the Higgs potential above the electroweak scale
\cite{Espinosa:2007qp,higgsstability}.

The conclusion however relies on the effective potential computed in
Minkowski space and one should ask if the effects of curvature can
be neglected during inflation. Indeed, treating the SM fields as
test fields in a de Sitter background, the
Higgs sector acquires a non-minimal coupling to gravity $\xi(\mu) R
\Phi^{\dag}\Phi$ through loop corrections even if $\xi=0$ at
tree-level \cite{Freedman:1974gs}. This effect and other curvature corrections could play
 a significant role under a high scale inflation where $R = 12 H^2$
 would be much larger than the (Minkowski) instability scale. In a tree-level analysis in
 \cite{Espinosa:2007qp} it was indeed found that
 the generated curvature mass stabilizes the SM vacuum during inflation for
 $\xi \gtrsim 10^{-1}$, however, loop corrections in the curved background
 space and the renormalization group (RG) running of the $\xi$-coupling were not addressed.

Ultimately, the quantity of interest is the transition rate or probability
 from the EW vacuum to the unstable false vacuum. In order to compute
 it reliably one should track the evolution of the Higgs fluctuations
 during inflation, which can be done for example by the stochastic Fokker-Planck (FP)
 equation (c.f. \cite{Hook:2014uia}). Clearly, curvature induced
 corrections by large IR fluctuations in de Sitter space are already
 accounted for by the FP equation, whereas the UV effects such as
 RG running of couplings are not reproduced, and hence must be
 incorporated directly in the input potential.

Therefore, it appears that a reliable approximation scheme would be to incorporate
 the UV-induced curvature corrections (arising from the UV part of the
 loop integrals) in the input effective potential, whereas the IR effects
 would be accounted for by the FP evolution equation itself.

In this work we compute the RG improved effective potential of the
 Standard Model Higgs in de Sitter space consistently accounting for the UV-induced
curvature corrections and their RG running with the measured best fit values of SM
 parameters at the EW scale as the input.
 We treat the SM fields as test fields in a fixed
inflationary background space assuming inflation is driven by new
physics not directly coupled to SM.
The renormalization procedure is
therefore not hampered by ambiguities of the gravitational sector as
for example in the Higgs inflation \cite{Bezrukov:2007ep}.
 We find that for a high inflationary scale the curvature corrections are generally
 significant and either stabilize or destabilize the potential against inflationary
 fluctuations, depending on the value of $\xi$-coupling at the EW scale.

Derivation of the effective action in curved background has been extensively studied in the
 literature \cite{Toms:1983qr}.
Here we use the resummed heat kernel expansion method \cite{Parker:1984dj} which incorporates
 complete UV contributions from the loop integrals at one-loop level. Unlike
 typically in the literature \cite{Toms:1983qr}, we do not treat the curvature scale $R$
 as a small expansion parameter, allowing us to consider the case where $R$ is larger than
 the tree-level masses of the standard model particles. Moreover, we fully incorporate
 RG improvement on top of the one-loop effective potential to lift the dependence on the
 renormalization scale, which is crucially important considering the large hierarchy
 between the EW and inflationary scales. This provides a significant improvement over the
 RG improved tree-level potential, for which the renormalization scale dependence is not
 cancelling up to one-loop level.


Using this method we find for the Standard Model one-loop effective potential in de Sitter
space improved by one-loop RG equations in the 't Hooft-Landau gauge and the $\overline{MS}$ scheme
\footnote{We use the notation of \cite{Casas:1994qy} with a modification $\lambda \to 2 \lambda$ according to
 more standard convention for the Higgs self-coupling.}
\ea{
V_{\rm eff} &= -\f{1}{2}m^2(t)\phi^2(t) + \f{1}{2}\xi(t)R\phi^2(t) + \f{1}{4}\lambda(t)\phi^4(t)
\nonumber\\
&+ \sum\limits_{i=1}^9 \f{n_i}{64\pi^2}M_i^4(\phi)\left[\log\f{\big|M_i^2(\phi)\big|}{\mu^2(t)} - c_i \right]\,,
\label{potential}
}
with
\ee{
M_i^2(\phi) = \kappa_i \phi^2(t) - \kappa'_i + \theta_i R\,,
\label{mass}
}
where the coefficients $n_i$, $\kappa_i$, $\kappa'_i$ and $\theta_i$ for various contributions are given
 by Table~\ref{tab:contributions}.
The parameters $\lambda(t)$ and $m(t)$ are the SM quartic
 coupling and mass, whereas $g(t)$, $g'(t)$ and $y_{\rm t}(t)$ are the SU(2), U(1) and top Yukawa couplings
 respectively, while $\xi(t)$ is the Higgs non-minimal coupling to gravity \footnote{Note that purely gravitational
 operators $R$, $R^2$ and $R^{\mu\nu}R_{\mu\nu}$ are required for the renormalization of the potential but are not
 directly relevant for our consideration.}. All of them are running with RGE.
 The running of the Higgs field is given by
\ee{
\phi(t) = Z(t)\phi_c\,,\qquad Z(t) = \exp\left[-\int_0^t dt' \gamma(t')\right]\,,
}
where $\phi_c$ is the classical field and $\gamma(t)$ is the Higgs field anomalous dimension. The scale
 $\mu(t)$ is related to the running parameter by
\ee{
\mu(t) = m_{\rm t} e^t\,,
}
where we have set the fixed scale at $t=0$ equal to physical top quark mass $m_{\rm t}$.
\begin{center}
\begin{table}
\caption{\label{tab:contributions}Contributions to the effective potential (\ref{potential})
 from $W^{\pm}$, $Z^0$, top quark t, Higgs $\phi$ and the Goldstone bosons $\chi_{1,2,3}$.}
\vspace{2mm}
 \begin{tabular}{|c||cccccc|}
 \hline
   $\Phi$ & $~~i$ & $~~n_i$  &$~~\kappa_i$ & $\kappa'_i$          & $~~\theta_i$   & $\quad c_i~~$ \\[1mm]\hline
   $~$ & $~~1$  & $~~2$       & $~~ g^2/4$        & $0$        & $~~{1}/{12}$     & $\quad{3}/{2}~~$ \\[1mm]
   $~W^\pm$ & $~~2$  & $~~6$       & $~~ g^2/4$        & $0$        & $~~{1}/{12}$      & $\quad{5}/{6}~~$ \\[1mm]
   $~$ & $~~3$  & $-2$      & $~~g^2/4$        & $0$        & $-{1}/{6}$      & $\quad{3}/{2}~~$ \\[1mm]\hline
   $~$ & $~~4$  & $~~1$       & $~~(g^2+g'^2)/4$ & $0$        & $~~{1}/{12}$     & $\quad{3}/{2}~~$ \\[1mm]
   $Z^0$ & $~~5$  & $~~3$       & $~~(g^2+g'^2)/4$ & $0$        & $~~{1}/{12}$      & $\quad{5}/{6}~~$ \\[1mm]
   $~$ & $~~6$  & $-1$      & $~~(g^2+g'^2)/4$ & $0$        & $-{1}/{6}$     & $\quad{3}/{2}~~$ \\[1mm]\hline
   t & $~~7$  & $-12$     & $~~ y_{\rm t}^2/2$      & $0$        & $~~{1}/{12}$     & $\quad{3}/{2}~~$ \\[1mm]\hline
   $\phi$ & $~~8$  & $~~1$       & $~~3\lambda$           & $~m^2$      & $~~\xi-{1}/{6}$  & $\quad{3}/{2}~~$
\\[1mm]\hline
   $\chi_i$ & $~~9$  & $~~3$       & $~~\lambda$            & $~m^2$      & $~~\xi-{1}/{6}$   & $\quad{3}/{2}~~$
\\[.5mm]\hline
  \end{tabular}
  \end{table}
\end{center}
A direct comparison with the flat space results \cite{Ford:1992mv,Casas:1994qy} shows that
 in this approximation spacetime curvature modifies the form of the effective potential
 only by shifting the effective masses by gravitational contributions proportional to $R$.
 We present more detailed derivation of the effective potential (\ref{potential}) elsewhere \cite{inpreparation}.
 For example, a generic gauge field contribution in arbitrary curved spacetime is given by
\ea{
V^{\rm (gauge)}_{\rm eff} =  \f{1}{64\pi^2}\bigg\{
&{\rm tr}\bigg[M_g^4\bigg(\log\f{|M_g^2|}{\tilde\mu^2} - \f{3}{2} \bigg)\bigg]
\nonumber\\&-M_s^4\left(\log\f{|M_s^2|}{\tilde\mu^2} - \f{3}{2} \right)\bigg\}\,,
\label{gauge_contr}
}
with ${\big(M_g^2\big)^\mu}_\nu = M_s^2 \delta^\mu_\nu + {R^\mu}_\nu$, $M_s^2 = m^2 - R/6$, where $m^2$ is the gauge boson
 mass term in the quadratic action and $\log(\tilde\mu^2) \equiv \log(4\pi\mu^2) + 2/(4-d) - \gamma_E$ contains the
 dimensional pole in the limit $d \to 4$. The relevant terms in the potential (\ref{potential}) are then found by
 computing the trace over spacetime indices using de Sitter space expression for the Ricci tensor:
\ee{
R = 12 H^2\,,
}
with ${R^\mu}_\nu = \delta^\mu_\nu R/4$.

\begin{figure}
\includegraphics[scale=0.8]{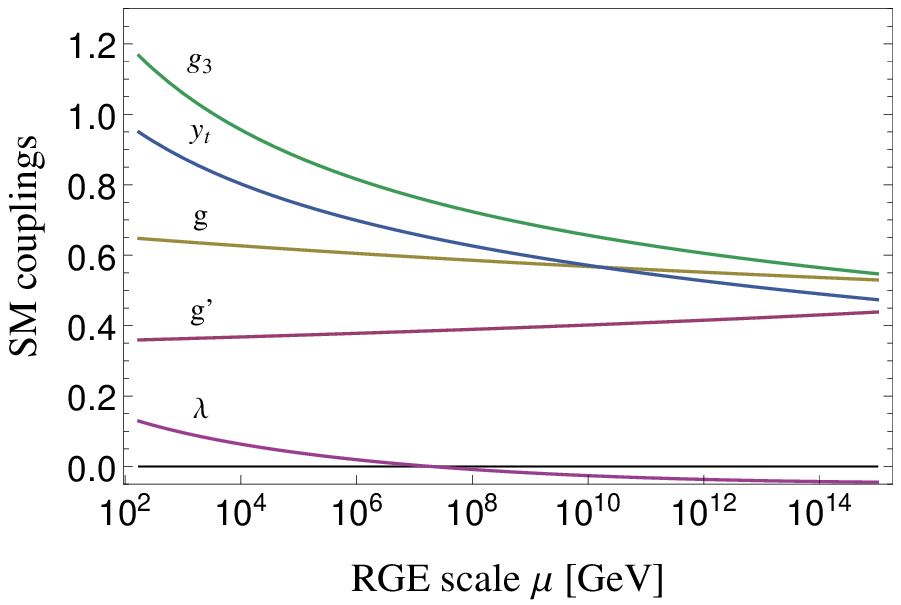}
\includegraphics[scale=0.8]{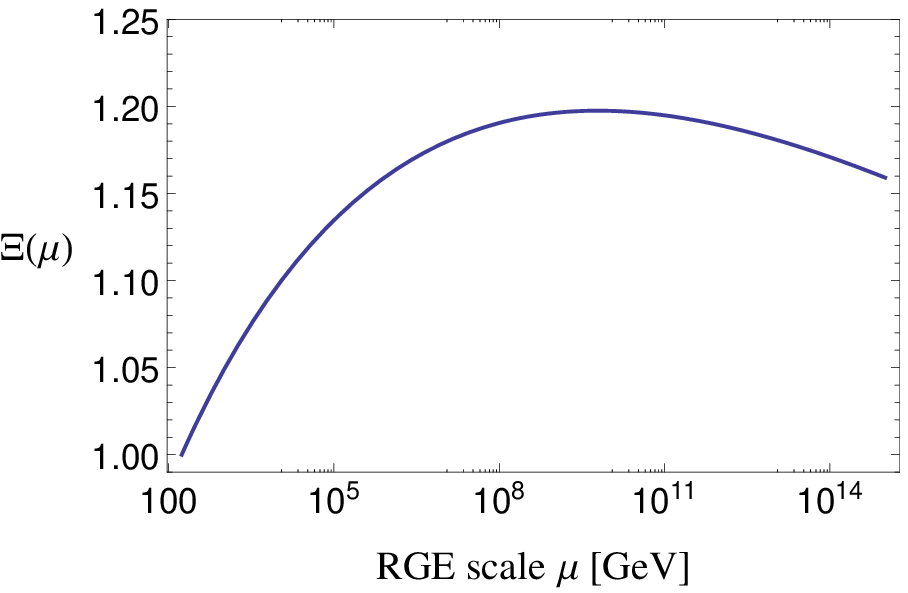}
\caption{
\label{fig:RGcouplings}
RGE running of the SM couplings $g_3$, $y_{\rm t}$, $g$, $g'$, $\lambda$ and the non-minimal gravity coupling $\Xi \equiv \big(\xi - 1/6 \big)/\big(\xi_{\rm EW} - 1/6 \big)$.
}
\end{figure}

The RG running is determined by the one-loop $\beta$- and $\gamma$-functions
 and the boundary conditions at the
 EW scale. At one-loop level the non-minimal gravity coupling $\xi$ does not couple into the $\beta$-functions of the
 SM couplings, which are therefore given by their usual expressions with NLO boundary conditions
 (c.f. \cite{Ford:1992mv,Buttazzo:2013uya}), resulting in the standard one-loop running as shown
 in Fig.~\ref{fig:RGcouplings}. The $\beta$-function for the non-minimal coupling $\xi$ scales as $\beta_{m^2}$ \cite{Buchbinder:1992rb} and is given by \footnote{The $\beta$-function for $\xi$ can be derived from the effective
 potential (\ref{potential}) by using the method of Coleman and Weinberg \cite{Coleman:1973jx}.}
\ee{
16\pi^2 \beta_{\xi} = \Big(\xi - \f{1}{6}\Big)\left(12\lambda + 6y_{\rm t}^2 - \f{3}{2}g'^2 - \f{9}{2}g^2\right)\,.
\label{betaxi}
}
It can be directly integrated by using the solutions for the running SM couplings to get
\ee{
\xi(t) = \f{1}{6} + \Big(\xi_{\rm EW} - \f{1}{6}\Big)\Xi(t)\,,
\label{running-xi}
}
where $\Xi(t)$ is shown in Fig.~\ref{fig:RGcouplings} and $\xi_{\rm EW} \equiv \xi(m_{\rm t})$ is the initial value
 at the electroweak scale.

The potential (\ref{potential}) is renormalization scale $\mu(t)$ invariant, if the
 derivative $dV_{\rm eff}/dt$ vanishes. By direct computation we get
\ea{
\f{d V_{\rm eff}}{dt} &=
\phi^4\left( \f{1}{4}\beta_\lambda - \gamma\lambda
 - \sum\limits_i \f{n_i \kappa_i^2}{32\pi^2}\right)\nonumber \\
&- \phi^2\Bigg( \f{1}{2}\beta_{m^2} - \gamma m^2
 - \sum\limits_i \f{n_i \kappa_i \kappa'_i}{16\pi^2}\Bigg)
\nonumber\\
&+ \phi^2 R \left( \f{1}{2}\beta_\xi - \gamma \xi
 - \sum\limits_i \f{n_i \kappa_i \theta_i}{16\pi^2}\right)
+ \ldots
\label{scale_invariance}
}
where the higher order contributions are neglected. Using the one-loop $\beta$- and $\gamma$-functions
 for the SM couplings (c.f. \cite{Buttazzo:2013uya}) and Eq.~(\ref{betaxi}) we explicitly find that each
 parenthesis in (\ref{scale_invariance}) vanishes and therefore the potential (\ref{potential}) is indeed
 renormalization scale invariant up to higher loop corrections.

The optimum scale $\mu_*$, where the neglected higher order corrections have the smallest impact on the
 observables, turns out to be certain average of the masses $M_i^2(\phi)$ such that the logarithms
 in (\ref{potential})
 do not result in large corrections. In the case of flat spacetime, $R=0$, it can be shown \cite{Ford:1992mv}
 that $\mu = \phi$ is a good choice resulting in small corrections to the optimal choice. Based on this
 consideration, we now make a choice of the scale $\mu$ in the presence of curvature corrections as
\ee{
\mu^2 = \phi^2 + R\,,
\label{scale_choice}
}
for which the corrections compared to the optimal choice are expected to be small \footnote{We have checked
 numerically that the corrections remain small if we choose $\mu^2 = \alpha\,\phi^2 + \beta\,R$
 and vary $\alpha$ and $\beta$ between $0.1$ and $10$.}.

\begin{figure}
\vspace{-0.2cm}
\includegraphics[scale=0.8]{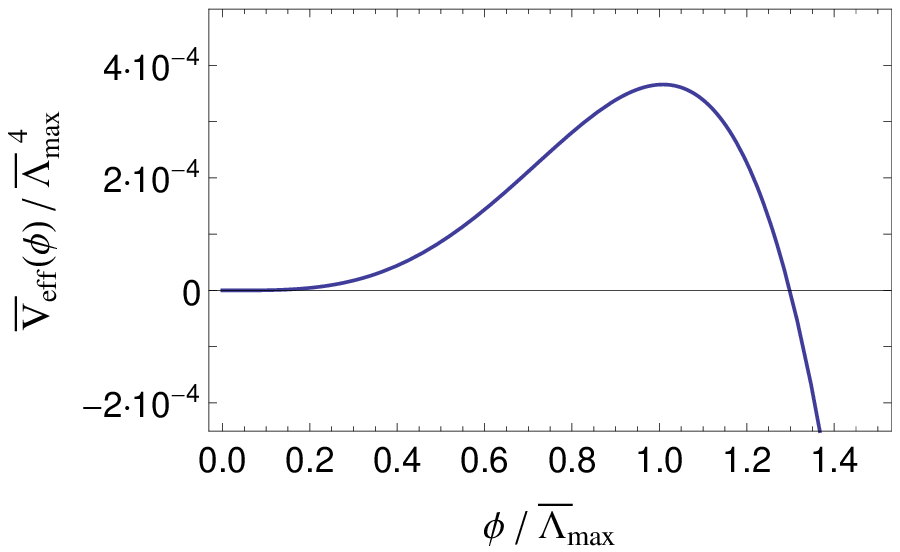}
\newline\newline \vspace{0.2cm}\hspace{-0.4cm}
\includegraphics[scale=0.8]{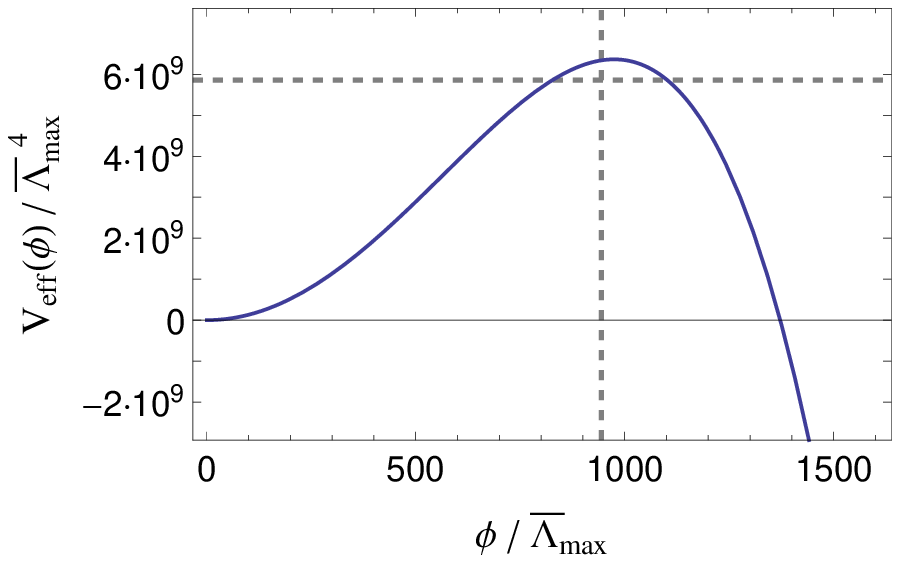}
\caption{
\label{fig:potential_plots}
The Higgs effective potential without (top) and with (bottom) curvature corrections for
 $H = 10^{10}\;{\rm GeV}$ and $\xi_{\rm EW} = 0.1$.
 The dashed lines correspond to the approximation (\ref{Lmax}-\ref{Vmax}) for the maximum.
}
\end{figure}

Once the running couplings are solved, the potential can be plotted by choosing the renormalization scale
 as in Eq.~(\ref{scale_choice}). The Minkowski potential, corresponding to $R = 0$, is shown in the upper panel of
 Fig.~\ref{fig:potential_plots}, where the scale of the maximum is given by
\ee{
\bar{\Lambda}_{\rm max} \simeq\,6 \cdot 10^{7}\;{\rm GeV}\,,\qquad
\bar{V}_{\rm max}^{1/4} \simeq\,9 \cdot 10^{6}\;{\rm GeV}\,,
}
where the bar on top of the symbol indicates quantities calculated from the Minkowski potential. With these
 values, inflationary fluctuations would be able to overcome the potential barrier if $H \gtrsim 10^7~{\rm GeV}$,
 rendering the physical vacuum unstable. At two-loop level the barrier is higher \cite{Degrassi:2012ry,Buttazzo:2013uya},
 $\bar{V}_{\rm max}^{1/4} \sim 10^9~{\rm GeV}$, but the instability remains although
 requiring a higher inflationary scale $H$.

The full effective potential
 (\ref{potential}) for a particular choice $H = 10^{10}\;{\rm GeV}$ and $\xi_{\rm EW} = 0.1$ of the free parameters,
 for which we find $\Lambda_{\rm max} \simeq 6 \cdot 10^{10}\;{\rm GeV}$ and
 $V_{\rm max}^{1/4} \simeq 2 \cdot 10^{10}\;{\rm GeV}$, is shown in the lower panel of Fig.~\ref{fig:potential_plots}.
 We find that the scale of the maximum is orders of magnitude higher than the prediction of the
 Minkowski potential due to large effective curvature mass.

\begin{figure}
\includegraphics[scale=0.8]{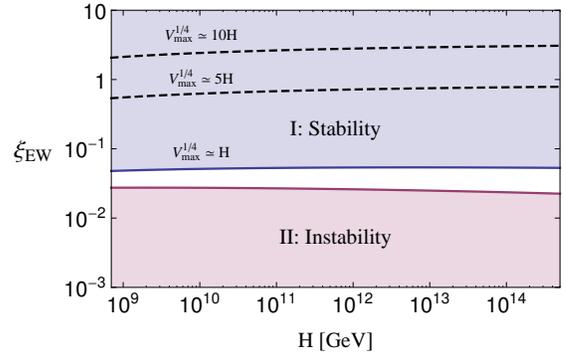}
\caption{The parametric regions I (blue, top), where $V_{\rm max}^{1/4} \gtrsim H$ and the transition probability
 to unstable vacuum is suppressed, and II (red, bottom), where the EW vacuum is unstable for all $\phi$ due to
 large negative curvature mass.}
\label{fig:regions}
\end{figure}

A reasonable order of magnitude estimate for the maximum of the potential at scales
 $H \gg \bar\Lambda_{\rm max} \sim 10^8\;{\rm GeV}$ can be obtained by fixing
 $\mu = R^{1/2} = 12^{1/2} H$, since the running couplings evolve mildly for $\mu \gtrsim 10^8\;{\rm GeV}$:
\ea{
\Lambda_{\rm max} &\simeq \left(\f{12\xi_R}{|\lambda_R|}\right)^{1/2} H
 \;\gtrsim\; \left(10^3 \xi_R\right)^{1/2} H\,,
\label{Lmax}
\\[2mm]
V_{\rm max}^{1/4} &\simeq \f{\left(6 \xi_R\right)^{1/2}}{|\lambda_R|^{1/4}} H\;\gtrsim\; 10\,\xi_R^{1/2}\,H\,,
\label{Vmax}
}
where we denote $\xi_R \equiv \xi\big(R^{1/2}\big)$ and similarly for $\lambda_R$ and
 we have used $|\lambda_R| \lesssim 10^{-2}$. Using Eqs.~(\ref{Vmax}) and (\ref{running-xi}) we then find that
 the criterion for stability:
\ee{
V_{\rm max}^{1/4} \gtrsim H\,,
}
can be solved for $\xi_{\rm EW}$ to get
\ee{
\xi_{\rm EW} \gtrsim \f{1}{6\,\Xi_R}\left(|\lambda_R|^{1/2} + \Xi_R - 1 \right) \sim 10^{-2}\,,
\label{stability_region}
}
where $\Xi_R \equiv \Xi\big(R^{1/2}\big) = 1.15 - 1.20$ for $H \gg 10^8\;{\rm GeV}$. We show the
 stability region given by Eq.~(\ref{stability_region}) as blue color in Fig.~\ref{fig:regions}.

\begin{figure}
\includegraphics[scale=0.8]{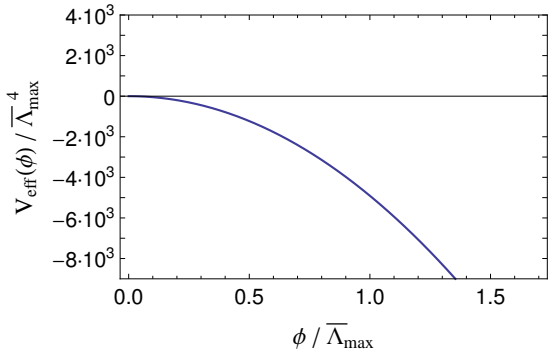}
\caption{The Higgs effective potential including curvature corrections with $H = 10^{10}\;{\rm GeV}$ and
 $\xi_{\rm EW} = 0$.}
\label{fig:potential_unstable}
\end{figure}

On the other hand, for small enough $\xi_{\rm EW}$ we find that $\xi(\mu)$ is running negative such that
 the negative curvature mass term $\xi R \phi^2/2$ dominates the potential (\ref{potential})
 at high scales if $H \gg \bar\Lambda_{\rm max}$. For example, for $\xi_{\rm EW} = 0$ the
 unstable potential is shown in Fig.~\ref{fig:potential_unstable}. The condition for $\xi_{\rm EW}$
 to yield unstable potential during inflation for $H \gg \bar\Lambda_{\rm max}$ is given by
 \ee{
\xi_{\rm EW} \lesssim \f{1}{6\,\Xi_I }\left(\Xi_I - 1 - \f{|\lambda_I|\bar\Lambda_{\rm max}^2}{4 H^2}\right)
 \sim 10^{-2}\,,
\label{instability_region}
}
where we denote $\Xi_I \equiv \Xi\big(\bar\Lambda_{\rm max}\big)$ and similarly for $\lambda_I$. The corresponding
 region where the EW vacuum is unstable from the onset of inflation is shown as red color in
 Fig.~\ref{fig:regions}.

The intermediate region between I and II in Fig.~\ref{fig:regions} is relatively narrow and requires
 fine-tuning for the non-minimal coupling $\xi$ at the EW scale. This is because $\xi$ is running away from the
 conformal point $\xi_c = 1/6$ by a factor of $1.15 - 1.20$ (see Fig.~\ref{fig:RGcouplings}), and therefore without
 fine-tuning $|\xi(\mu)|$ is typically of order $10^{-2}$ or larger at high scales. The curvature corrections may
 be important in this region as well, however, further investigation is required for a conclusive survey.

In conclusion, we find that for a high inflationary scale $H \gg \bar\Lambda_{\rm max} \sim 10^8\;{\rm GeV}$
 the UV-induced (subhorizon) curvature corrections alter the SM Higgs effective potential significantly during
 inflation. In particular, for $\xi_{\rm EW} \gtrsim 6\cdot 10^{-2}$ a large curvature mass stabilizes the
 potential against fluctuations of order $H$, while for $\xi_{\rm EW} \lesssim 2 \cdot 10^{-2}$ the resulting
 curvature mass is negative such that the EW vacuum is unstable from the onset of
 inflation. These results are in agreement with the tree-level analysis in \cite{Espinosa:2007qp} where the
 stability bound was found to be $\xi \gtrsim 10^{-1}$.
 We will examine the vacuum transition rate and the implications on
 cosmology in more detail elsewhere \cite{inpreparation}. We stress
 however that the exponential suppression of any fluctuation probability $P\sim
 {\rm exp}(-V_{\rm max}/H^4)$ makes the stability of the regime $V_{\rm max}^{1/4} \gtrsim
 H$ a robust statement.

Finally, we note that higher loop corrections may alter these
quantitative estimates considerably. For example, the flat space
instability scale $\bar{\Lambda}_{\rm max} \sim 10^{\rm 11}~{\rm
GeV}$ from the NNLO calculation
\cite{Degrassi:2012ry,Buttazzo:2013uya} is roughly three orders of
magnitude higher
 than in the present one-loop calculation. However, we expect that our qualitative results persist.

\acknowledgments{MH and TM would like to thank Anders Tranberg for useful and illuminating discussions and the
 University of Stavanger for hospitality. MH is supported by the Villum Foundation Grant No. YIP/VKR022599,
 TM and SN are supported by the Academy of Finland Grants No. 1134018 and 257532, respectively,
 and AR is supported by STFC Grant No. ST/J000353/1.}

\end{document}